\documentclass[12pt]{article}
\newcommand{\xrm}[1]{{\textstyle \mbox{\rm #1}}}
\newcommand{\bm}[1]{\mbox{\boldmath $#1$}}
\newcommand{\fnd}[2]{\frac{\textstyle #1}{\textstyle #2}}
\newcommand{\abs}[1]{\left| #1\right|}
\newcommand{\Real}[1]{\Re {\it e}(#1 )}
\newcommand{\Imag}[1]{\Im {\it m}(#1 )}

\begin{document}
\title{\bf Behaviour of \bm{S}-Wave poles Near Threshold
and the Scalar Meson Nonet Below 1 GeV}
\author{{\large Eef van Beveren{\normalsize $^{\; a}$} and
George Rupp{\normalsize $^{\; b}$}}\\ [.5cm]
{\normalsize\it $^{a\;}$Centro de F\'{\i}sica Te\'{o}rica, Departamento de
F\'{\i}sica,}\\ {\normalsize\it Universidade, P3004-516 Coimbra, Portugal,}
{\small (eef@teor.fis.uc.pt)}\\ [.3cm]
{\normalsize\it $^{b\;}$Centro de F\'{\i}sica das Interac\c{c}\~{o}es
Fundamentais, Instituto Superior T\'{e}cnico,}\\ {\normalsize\it
Edif\'{\i}cio Ci\^{e}ncia,
P1049-001 Lisboa Codex, Portugal,} {\small (george@ajax.ist.utl.pt)}
}
\maketitle

\begin{abstract}
We describe the behaviour at threshold of $S$-wave poles of the
scattering matrix within a four-parameter model for non-exotic
meson-meson scattering. This model accommodates all non-exotic mesons,
hence also the light scalar mesons, as resonances and bound states
characterised by complex singularities of the scattering amplitude
as a function of the total invariant mass.

The majority of the full $S$-matrix mesonic poles
stem from an underlying confinement spectrum.
However, the light scalar mesons $K_{0}^{\ast}(830)$, $a_{0}(980)$,
$f_{0}$(400--1200), and $f_{0}(980)$ do not,
but instead originate in $^{3}P_{0}$-barrier semi-bound states.
We show that the behaviour of the corresponding poles is
identical at threshold.
\end{abstract}

\section*{Introduction}

In a theory with quarks and mesons one can study strong interactions
through meson-meson scattering.
Observed spectra may be interpreted in terms of quark-antiquark
or more complicated systems \cite{HEPPH0206263}.
Necessarily, spectra must be confirmed in several different
experiments, and compiled in spectroscopic tables.
Then, for a lowest-order approximation of strong interactions,
confinement models may be constructed.
Their usefulness can be measured by the model's achievements
when adjusting its parameters to experiment\cite{PRD32p189}.

For further refinements of strong-interaction models, there
are several possible directions.
One way is to reanalyse experiments in order to adjust the data
to new insights \cite{ZPC51p689}, in particular when popular models
\cite{CERN-TH-5246-88} are not capable of reproducing specific experimental
results \cite{NPPS21p105,HADRON91p410,SLAC-PUB-5606,SLAC-PUB-5657}.
Another way is to adapt the model by including more sophisticated
interactions \cite{PLB409p483,EPJA9p221},
though often at the cost of more model parameters.
One may also just ignore the spectroscopy which is distilled from
experiment, and, instead of comparing resonance positions and
widths to information from the spectroscopic tables,
compare the model's predictions directly, if possible, to experimental
scattering cross sections and phase shifts
\cite{Cargese75p305,PRD21p203,PRD27p1527,ZPC30p615,NPB266p451,PRL76p1575,HEPPH0110081,HEPPH0203255}.

Here we will discuss the eternally disputed
\cite{HEPPH0204205,HEPPH0201171}
low-lying nonet of $S$-wave poles in meson-meson scattering cross sections,
which are predicted by a four-parameter model \cite{ZPC30p615}.
It should thereby be noted that the model's parameters have already
been fitted to the $J^{P}=1^{-}$ $c\bar{c}$ and $b\bar{b}$ spectra,
as well as to $P$-wave meson-meson scattering data \cite{PRD27p1527}.

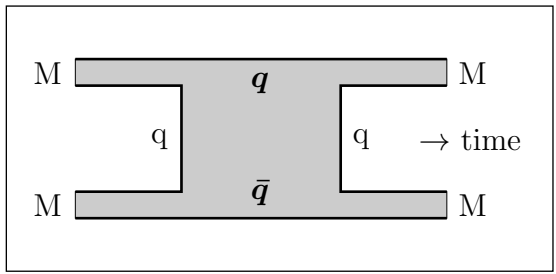
\begin{figure}[ht]
\begin{center}
\begin{picture}(206,100)(-6,0)
\put(15,25){\makebox(0,0)[rc]{M}}
\put(15,75){\makebox(0,0)[rc]{M}}
\put(165,25){\makebox(0,0)[lc]{M}}
\put(165,75){\makebox(0,0)[lc]{M}}
\put(90,25){\makebox(0,0)[bc]{\bm{\bar{q}}}}
\put(90,75){\makebox(0,0)[tc]{\bm{q}}}
\put(55,50){\makebox(0,0)[rc]{q}}
\put(125,50){\makebox(0,0)[lc]{q}}
\put(150,50){\makebox(0,0)[lc]{$\rightarrow$ time}}
\end{picture}
\end{center}
\caption[]{The mesons (M) exchange a quark or, equivalently,
a quark-antiquark pair is annihilated,
followed by a second quark exchange, equivalent
to a new quark-antiquark pair being created.}
\label{qexch}
\end{figure}

\begin{figure}[ht]
\begin{center}
\begin{picture}(206,100)(-6,0)
\put(15,60){\makebox(0,0)[rc]{M}}
\put(185,60){\makebox(0,0)[lc]{M}}
\put(100,35){\makebox(0,0)[lt]{meson loop}}
\end{picture}
\end{center}
\caption[]{The lowest-order self-energy graph for a meson propagator.}
\label{bubble}
\end{figure}
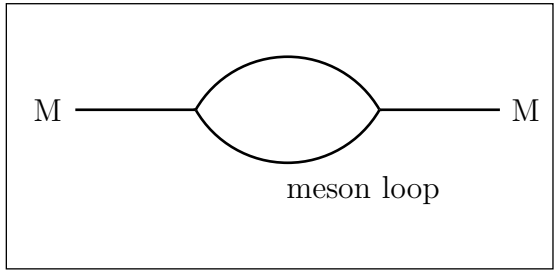

First, let us outline the motivation for our work.
For the interaction in the vicinity of a resonance in meson-meson
scattering, one may consider quark-exchange
or quark-pair-creation processes, as depicted in Fig.~(\ref{qexch}).
When the intermediate $q\bar{q}$ system is close enough to a genuine
bound state of confinement, then the system will resonate,
resulting in a resonance in meson-meson scattering.
Another picture for the same phenomenon, shown in Fig.~(\ref{bubble}),
is that a mesonic quark-antiquark system (M) gets a self-energy correction
from a virtual meson loop.

Either picture describes the same physical situation,
namely a mesonic resonance or bound state \cite{PLB509p81},
but in a rather different way.
Our aim is to merge both pictures in one model.

\section{Quark exchange}
\label{QExchange}

In the quark-exchange picture we obtain a resonance in the particular
partial-wave meson-meson-scattering cross section which agrees with
the quantum numbers of the intermediate $q\bar{q}$ system.

Such a process may be described by scattering phase shifts of the form

\begin{equation}
\xrm{cotg}\left(\delta_{\ell}(s)\right)\;\approx\;
\fnd{E_{R}-\sqrt{s}}{\Gamma_{R}/2}
\;\;\; ,
\label{cotgdR}
\end{equation}

\noindent
where $E_{R}$ and $\Gamma_{R}$ represent the central
invariant meson-meson mass and the resonance width, respectively.

However, formula (\ref{cotgdR}) is a good approximation for the
scattering cross section only when the resonance shape is not very much
distorted and the width of the resonance is small.
Moreover, the intermediate state in such a process is essentially
a constituent $q\bar{q}$ configuration that is part of a
confinement spectrum (also referred to as bare or intrinsic states),
and hence may resonate in one of the eigenstates.
This implies that the colliding mesons scatter off
the whole $q\bar{q}$ confinement spectrum of radial, and possibly
also angular excitations, not just off one single state \cite{NC14p951}.
Consequently, a full expression for the phase shifts of formula
(\ref{cotgdR}) should contain all possible eigenstates of such a
spectrum as long as quantum numbers are respected.
Let us denote the eigenvalues of the relevant part of the spectrum
by $E_{n}$ ($n=0$, $1$, $2$, $\dots$)
and the corresponding eigenstates by $F_{n}$.
Then, following the procedure outlined
in Ref.~\cite{HEPEX0106077},
we may write for the partial-wave phase shifts the
more general expression

\begin{equation}
\xrm{cotg}\left(\delta (s)\right)\; =\;
\left[ I(s)\;\sum_{n=0}^{\infty}\fnd{
\abs{{\cal F}_{n}}^{2}}{\sqrt{s}-E_{n}}\right]^{-1}\;
\left[ R(s)\;\sum_{n=0}^{\infty}\fnd{
\abs{{\cal F}_{n}}^{2}}{\sqrt{s}-E_{n}}\; -\; 1\right]
\;\;\; .
\label{cotgdS}
\end{equation}

\noindent
In $R(s)$ and $I(s)$ we have absorbed the kinematical factors
and the details of two-meson scattering, and
moreover the three-meson vertices.
The details of formula (\ref{cotgdS}) can be found in
Ref.~\cite{HEPEX0106077}.

For an approximate description of a specific resonance and in
the rather hypothetical case that the three-meson vertices have small
coupling constants, one may single out, from the sum over all
confinement states, one particular state (say number $N$),
the eigenvalue of which is nearest to the
invariant meson-meson mass close to the resonance.
Then, for total invariant meson-meson masses $\sqrt{s}$ in the vicinity
of $E_{N}$, one finds the approximation

\begin{equation}
\xrm{cotg}\left(\delta (s)\right)\;\approx\;
\fnd{\left[ E_{N}\; +\; R(s)\;\abs{{\cal F}_{N}}^{2}\right]\; -\;\sqrt{s}}
{I(s)\;\abs{{\cal F}_{N}}^{2}}
\;\;\; .
\label{cotgdSs}
\end{equation}

\noindent
Formula (\ref{cotgdSs}) is indeed of the form (\ref{cotgdR}),
with the central resonance position and width given by

\begin{equation}
E_{R}\;\approx\; E_{N}\; +\; R(s)\;\abs{{\cal F}_{N}}^{2}
\;\;\;\;\;\xrm{and}\;\;\;\;\;
\Gamma_{R}\;\approx\; 2I(s)\;\abs{{\cal F}_{N}}^{2}
\;\;\; .
\label{ERGR}
\end{equation}

\noindent
In experiment one observes the influence of the nearest bound state of
the confinement spectrum, as in classical resonating systems.
Nevertheless, formula (\ref{cotgdSs}) is only a good approximation
when the three-meson couplings are small.
Since the coupling of the meson-meson system to quark exchange
is strong, the influence of the higher- and lower-lying excitations is
not negligible.

In the other hypothetical limit, namely of very large couplings, we obtain
for the phase shift the expression

\begin{equation}
\xrm{cotg}\left(\delta (s)\right)\;\approx\;
\fnd{R(s)}{I(s)}
\;\;\; ,
\label{cotgdSl}
\end{equation}

\noindent
which describes scattering off an infinitely hard cavity.

The physical values of the couplings come out somewhere in between
the two hypothetical cases.
Most resonances and bound states can be classified as stemming from a
specific confinement state \cite{PLB413p137,HEPPH0204328}.
However, some structures in the scattering cross section stem from the
cavity which is formed by quark exchange or pair creation
\cite{HEPEX0106077}.
The most notable of such states are the low-lying resonances
observed in $S$-wave pseudoscalar-pseudoscalar scattering
\cite{HEPEX0012009,HEPEX0110052,HEPEX0204018,HEPPH0110156}.

\section{Meson loops}

>From the discussion in Sec.~(\ref{QExchange}) one may conclude
that to lowest order the mass of a meson follows from
the quark-antiquark confinement spectrum.
It is however well-known that higher-order contributions
to the meson propagator, in particular those from meson loops,
cannot be neglected.

Virtual meson loops give a correction to the meson mass, whereas decay
channels also contribute to the strong width of the meson.
One obtains for the propagator of a meson the form

\begin{equation}
\Pi(s)\;= \;\fnd{1}
{s\; -\;\left( M_\xrm{confinement}\; +\;\sum\;\Delta M_\xrm{meson loops}
\right)^{2}}
\;\;\; ,
\label{propag}
\end{equation}

\noindent
where $\Delta M$ develops complex values when open decay channels are involved.

For the full mass of a meson, all possible meson-meson loops have to be
considered.
A model for meson-meson scattering must therefore include all
possible inelastic channels as well.
Although in principle this could be done, in practice it is not manageable,
unless a scheme exists dealing with all vertices and their relative
intensities.
In Ref.~\cite{ZPC21p291} relative couplings have been determined in the
harmonic-oscillator approximation assuming $^{3}P_{0}$ quark exchange.
However, further kinematical factors must be worked out and included.

\section{The spectrum}

The complete model consists of an expression for the $K$ matrix,
similar to formula (\ref{cotgdS}),
but extended to many meson-meson scattering channels,
several constituent quark-antiquark channels, and more
complicated transition potentials \cite{PRD27p1527,ZPC30p615},
which at the same time
and with the same set of four parameters reproduces bound states,
partial-wave scattering quantities, and the electromagnetic
transitions of $c\bar{c}$ and $b\bar{b}$ systems \cite{PRD44p2803}.

The $K$ matrix can be analytically continued below the various
thresholds, even the lowest one,
with no need of redefining any of the functions involved,
in order to study the singularities of the corresponding scattering matrix.
Below the lowest threshold, these poles show up on the real $\sqrt{s}$
axis, and can be interpreted as the bound states of the coupled system, to be
identified with the stable mesons.
For the light flavours one finds this way a nonet of light
pseudoscalars, i.e., the pion, Kaon, eta, and eta$'$.
For the heavy flavours, the lowest-lying model poles
can be identified with the $D(1870)$, $D_{s}(1970)$, $\eta_{c}(1S)$,
$J/\psi (1S)$, $\psi (3686)$, $B(5280)$, $B_{s}(5380)$,
$\Upsilon (1S)$, $\Upsilon (2S)$, and $\Upsilon (3S)$.

Above the lowest threshold, the model's partial-wave cross sections
and phase shifts for all included meson-meson channels
can be calculated and compared to experiment, as well as the
inelastic transitions.
Singularities of the scattering matrix come out with negative imaginary
part in the $\sqrt{s}$ plane.
To say it more precisely: out of the many singularities in a rather complex
set of Riemann sheets, some come close enough to the physical real
axis to be noticed in the partial-wave phase shifts and
cross sections.
In fact, each meson-meson channel doubles the number of Riemann
sheets, hence the number of poles.
Consequently, with ten scattering channels one has for each eigenvalue of
the confinement spectrum 1024 poles in 1024 Riemann sheets, out
of which usually only one contains relevant poles in each $\sqrt{s}$
interval in between the thresholds.
Those can be identified with the known resonances,
like the $\rho$ pole in $\pi\pi$ scattering,
or the $K^{\ast}$ pole in $K\pi$ scattering.
However, there might always be a pole in a nearby Riemann sheet
just around the corner of one of the thresholds, which can be
noticed in the partial-wave cross section.
The study of poles is an interesting subject by itself
\cite{NPB587p331,PRD59p074001}.

Once the four model parameters have been adjusted to
the experimental phase shifts and cross sections,
the pole positions can be determined and compared to tables for
meson spectroscopy.
But usually no new information is gained from such a comparison.
Here we will report on the singularity structure of the lowest
poles in $S$-wave meson-meson scattering.

\section{Scattering-matrix poles}
\label{scatpoles}

In the hypothetical case of very small couplings for the three-meson
vertices, we obtain poles in the scattering matrix that are close
to the eigenvalues of the confinement spectrum.
Let us denote by $M_{1}$ and $M_{2}$ the meson masses,
and by $\Delta E$ the difference between the complex-energy pole
of the scattering matrix and the energy eigenvalue, $E_{N}$, of
the nearby state of the confinement spectrum.
Using formula (\ref{ERGR}), we obtain

\begin{equation}
\Delta E\;\approx\;
\left\{ R(s)\; -\; iI(s)\right\}\;\abs{{\cal F}_{N}}^{2}
\;\;\; .
\label{DeltaE}
\end{equation}

\noindent
We may distinguish two different cases:
\vspace{0.3cm}

(1) $E_{N}\;>\; M_{1}+M_{2}$ (above threshold),
\vspace{0.1cm}

(2) $E_{N}\;<\; M_{1}+M_{2}$ (below threshold).
\vspace{0.5cm}

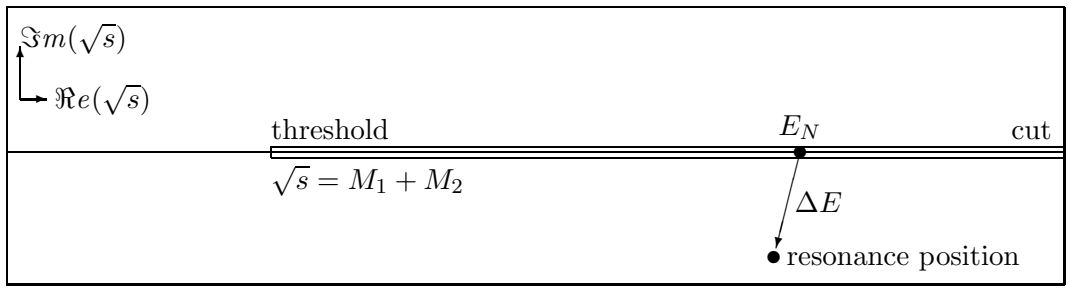
\begin{figure}[ht]
\begin{center}
\begin{picture}(400,105)(0,-50)
\put(  0,-50){\line(1,0){400}}
\put(  0, 55){\line(1,0){400}}
\put(  0,-50){\line(0,1){105}}
\put(400,-50){\line(0,1){105}}
\put(0,0){\line(1,0){400}}
\put(100,-2){\line(0,1){4}}
\put(100,-2){\line(1,0){300}}
\put(100, 2){\line(1,0){300}}
\put(380,5){\makebox(0,0)[bl]{\small cut}}
\put(100,5){\makebox(0,0)[bl]{\small threshold}}
\put(100,-5){\makebox(0,0)[tl]{\small $\sqrt{s}=M_{1}+M_{2}$}}
\put(300,0){\makebox(0,0){$\bullet$}}
\put(300,5){\makebox(0,0)[bc]{\small $E_{N}$}}
\put(300,0){\vector(-1,-4){9}}
\put(298,-15){\makebox(0,0)[lt]{\small $\Delta E$}}
\put(290,-40){\makebox(0,0){$\bullet$}}
\put(295,-40){\makebox(0,0)[lc]{\small resonance position}}
\put(5,20){\vector(0,1){20}}
\put(5,43){\makebox(0,0)[cl]{\small $\Imag{\sqrt{s}}$}}
\put(5,20){\vector(1,0){10}}
\put(18,20){\makebox(0,0)[cl]{\small $\Real{\sqrt{s}}$}}
\end{picture}
\end{center}
\caption[]{When the confinement state on the real $\sqrt{s}$ axis is in
the scattering continuum, then for small coupling (perturbative regime)
the resonance pole moves into the lower half of the complex $\sqrt{s}$ plane.}
\label{Above}
\end{figure}

When the nearby state of the confinement spectrum is in the
scattering continuum, then $\Delta E$ has a {\bf negative}
imaginary part and a real part, since both $R(s)$ and $I(s)$
of formula (\ref{DeltaE}) are real, and $I(s)$ is moreover positive.
The resonance singularity of the scattering matrix which corresponds to
this situation is depicted in Fig.~(\ref{Above}).

Notice that the resonance singularity is in the lower half of the
complex-energy plane (second Riemann sheet), as it should be.

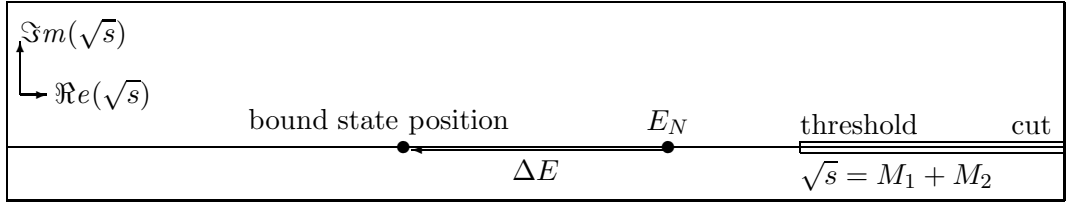
\begin{figure}[ht]
\begin{center}
\begin{picture}(400,75)(0,-20)
\put(  0,-20){\line(1,0){400}}
\put(  0, 55){\line(1,0){400}}
\put(  0,-20){\line(0,1){75}}
\put(400,-20){\line(0,1){75}}
\put(0,0){\line(1,0){400}}
\put(300,-2){\line(0,1){4}}
\put(300,-2){\line(1,0){100}}
\put(300, 2){\line(1,0){100}}
\put(380,5){\makebox(0,0)[bl]{\small cut}}
\put(300,5){\makebox(0,0)[bl]{\small threshold}}
\put(300,-5){\makebox(0,0)[tl]{\small $\sqrt{s}=M_{1}+M_{2}$}}
\put(250,0){\makebox(0,0){$\bullet$}}
\put(250,5){\makebox(0,0)[bc]{\small $E_{N}$}}
\put(250,-1){\vector(-1,0){97}}
\put(200,-5){\makebox(0,0)[tc]{\small $\Delta E$}}
\put(150,0){\makebox(0,0){$\bullet$}}
\put(190,5){\makebox(0,0)[rb]{\small bound state position}}
\put(5,20){\vector(0,1){20}}
\put(5,43){\makebox(0,0)[cl]{\small $\Imag{\sqrt{s}}$}}
\put(5,20){\vector(1,0){10}}
\put(18,20){\makebox(0,0)[cl]{\small $\Real{\sqrt{s}}$}}
\end{picture}
\end{center}
\caption[]{When the confinement state on the real $\sqrt{s}$ axis is
below the lowest scattering threshold, then the bound-state singularity
comes out on the real $\sqrt{s}$ axis.}
\label{Below}
\end{figure}

When the nearby state of the confinement spectrum is below
the scattering threshold, then $\Delta E$ has only a real part,
since $I(s)$ turns purely imaginary below threshold, whereas
$R(s)$ remains real.
The bound-state singularity of the scattering matrix corresponding to
this situation is depicted in Fig.~(\ref{Below}).

Note that the bound-state pole is on the real axis of
the complex-energy plane, as it should be.

\section{Threshold behaviour}

Near the lowest threshold, as a function of the overall coupling
constant, $S$-wave poles behave very differently
from $P$- and higher-wave poles.
This can easily be understood from the effective-range expansion
\cite{PotentialScattering} at the pole position.
There, the cotangent of the phase shift equals $i$.
Hence, for $S$ waves the next-to-lowest-order term in the expansion
equals $ik$ ($k$ represents the linear momentum related to $s$
and to the lowest threshold).
For higher waves, on the other hand, the next-to-lowest-order term in the
effective-range expansion is proportional to $k^{2}$.

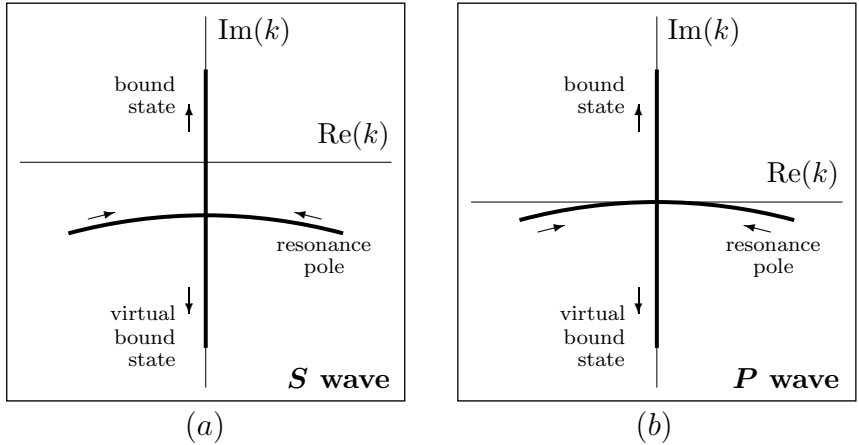
\begin{figure}[ht]
\begin{center}
\begin{picture}(320,160)(0,-10)
\put(145,95){\makebox(0,0)[rb]{\small Re($k$)}}
\put(80,145){\makebox(0,0)[lt]{\small Im($k$)}}
\put(31,69){\vector(4,1){10}}
\put(119,69){\vector(-4,1){10}}
\put(69,102){\vector(0,1){10}}
\put(69,43){\vector(0,-1){10}}
\put(64,36){\makebox(0,0)[rt]{\scriptsize virtual}}
\put(64,27){\makebox(0,0)[rt]{\scriptsize bound}}
\put(64,18){\makebox(0,0)[rt]{\scriptsize state}}
\put(64,123){\makebox(0,0)[rt]{\scriptsize bound}}
\put(64,114){\makebox(0,0)[rt]{\scriptsize state}}
\put(120,60){\makebox(0,0)[ct]{\scriptsize resonance}}
\put(120,54){\makebox(0,0)[ct]{\scriptsize pole}}
\put(75,-4){\makebox(0,0)[ct]{$(a)$}}
\put(145,5){\makebox(0,0)[rb]{\small\bf \bm{S} wave}}
\put(315,80){\makebox(0,0)[rb]{\small Re($k$)}}
\put(250,145){\makebox(0,0)[lt]{\small Im($k$)}}
\put(201,64){\vector(4,1){10}}
\put(289,64){\vector(-4,1){10}}
\put(239,102){\vector(0,1){10}}
\put(239,43){\vector(0,-1){10}}
\put(234,36){\makebox(0,0)[rt]{\scriptsize virtual}}
\put(234,27){\makebox(0,0)[rt]{\scriptsize bound}}
\put(234,18){\makebox(0,0)[rt]{\scriptsize state}}
\put(234,123){\makebox(0,0)[rt]{\scriptsize bound}}
\put(234,114){\makebox(0,0)[rt]{\scriptsize state}}
\put(290,60){\makebox(0,0)[ct]{\scriptsize resonance}}
\put(290,54){\makebox(0,0)[ct]{\scriptsize pole}}
\put(245,-4){\makebox(0,0)[ct]{$(b)$}}
\put(315,5){\makebox(0,0)[rb]{\small\bf \bm{P} wave}}
\end{picture}
\end{center}
\caption[]{Variation of the positions of scattering-matrix poles
as a function of hypothetical variations in the three-meson-vertex coupling,
for $S$ waves ($a$), and for $P$ and higher waves ($b$).
The arrows indicate increasing coupling constant.}
\label{SPDpoles}
\end{figure}

Poles for $P$ and higher waves behave in the complex $k$ plane
as indicated in Fig.~(\ref{SPDpoles}$b$).
The two $k$-plane poles meet at threshold ($k=0$).
When the coupling constant of the model is increased, the poles
move along the imaginary $k$ axis.
One pole moves towards negative imaginary $k$, corresponding to
a virtual bound state below threshold on the real $\sqrt{s}$ axis,
but in the wrong Riemann sheet.
The other pole moves towards positive imaginary $k$,
corresponding to a real bound state.

For $S$-wave poles, the behaviour is shown in Fig.~(\ref{SPDpoles}$a$).
The two $k$-plane poles meet on the negative imaginary $k$ axis.
When the coupling constant of the model is slightly increased,
both poles continue on the negative imaginary $k$ axis,
corresponding to two virtual bound states below threshold on the real
$\sqrt{s}$ axis.
Upon further increasing the coupling constant of the model,
one pole moves towards increasing negative imaginary $k$, thereby
remaining a virtual bound state for all values of the coupling constant.
The other pole moves towards positive imaginary $k$,
eventually passing threshold ($k=0$), thereby turning into a real
bound state of the system of coupled meson-meson scattering channels.
Hence, for a small range of hypothetical values of the coupling constant,
there are two virtual bound states, one of which is very close to
threshold.
Such a pole certainly has noticeable influence on the scattering cross section.

Although we are not aware of any experimental data that could
confirm the above-described threshold behaviour of poles,
we suspect this to be possible for atomic transitions in cavities.
Unfortunately, it does not seem likely that in the near future
similar processes can be studied for strong coupling.

\section{The low-lying nonet of \bm{S}-wave poles}

The nonet of low-lying $S$-wave poles behave as described
in Sec.~(\ref{scatpoles}), with respect to variations of the
model's overall coupling constant.
However, they do not stem from the confinement spectrum,
but rather from the cavity.
For small values of the coupling, such poles disappear into the
continuum, i.e., they move towards negative imaginary infinity
\cite{HEPEX0106077},
and not towards an eigenstate of the confinement spectrum as in
Fig.~(\ref{Above}).

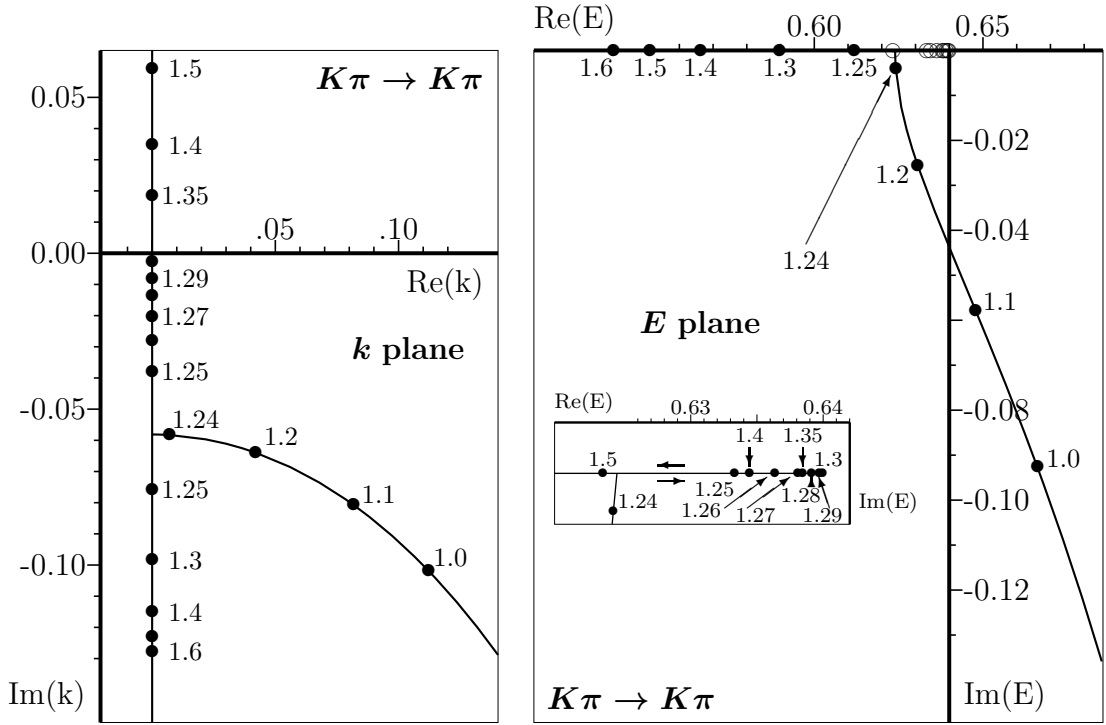
\begin{figure}[ht]
\normalsize
\begin{center}
\begin{picture}(190,280)(-30,0)
\put(66.11,183.15){\makebox(0,0)[bc]{.05}}
\put(112.71,183.15){\makebox(0,0)[bc]{.10}}
\put(-5.52,59.61){\makebox(0,0)[rc]{-0.10}}
\put(-5.52,118.62){\makebox(0,0)[rc]{-0.05}}
\put(-5.52,177.63){\makebox(0,0)[rc]{0.00}}
\put(-5.52,236.64){\makebox(0,0)[rc]{0.05}}
\put(145,172){\makebox(0,0)[tr]{Re(k)}}
\put(-5.52,4.01){\makebox(0,0)[br]{Im(k)}}
\put(124.07,57.41){\makebox(0,0){$\bullet$}}
\put(95.66,82.38){\makebox(0,0){$\bullet$}}
\put(58.69,102.13){\makebox(0,0){$\bullet$}}
\put(26.13,108.83){\makebox(0,0){$\bullet$}}
\put(19.53,88.07){\makebox(0,0){$\bullet$}}
\put(19.53,61.34){\makebox(0,0){$\bullet$}}
\put(19.53,41.77){\makebox(0,0){$\bullet$}}
\put(19.53,32.35){\makebox(0,0){$\bullet$}}
\put(19.53,26.76){\makebox(0,0){$\bullet$}}
\put(19.53,247.47){\makebox(0,0){$\bullet$}}
\put(19.53,218.55){\makebox(0,0){$\bullet$}}
\put(19.53,199.28){\makebox(0,0){$\bullet$}}
\put(19.53,174.22){\makebox(0,0){$\bullet$}}
\put(19.53,168.04){\makebox(0,0){$\bullet$}}
\put(19.53,161.23){\makebox(0,0){$\bullet$}}
\put(19.53,153.53){\makebox(0,0){$\bullet$}}
\put(19.53,144.48){\makebox(0,0){$\bullet$}}
\put(19.53,132.76){\makebox(0,0){$\bullet$}}
\footnotesize
\put(132.40,62.41){\makebox(0,0){1.0}}
\put(105.23,86.38){\makebox(0,0){1.1}}
\put(68.26,108.13){\makebox(0,0){1.2}}
\put(35.99,114.83){\makebox(0,0){1.24}}
\put(32,88.07){\makebox(0,0){1.25}}
\put(32,61.34){\makebox(0,0){1.3}}
\put(32,41.77){\makebox(0,0){1.4}}
\put(32,26.76){\makebox(0,0){1.6}}
\put(32,247.47){\makebox(0,0){1.5}}
\put(32,218.55){\makebox(0,0){1.4}}
\put(32,199.28){\makebox(0,0){1.35}}
\put(32,168.04){\makebox(0,0){1.29}}
\put(32,153.53){\makebox(0,0){1.27}}
\put(32,132.76){\makebox(0,0){1.25}}
\normalsize
\put(145,248){\makebox(0,0)[tr]{\bm{K\pi\rightarrow K\pi}}}
\put(95,135){\makebox(0,0)[bl]{\bm{k} {\bf plane}}}
\end{picture}
\begin{picture}(215,280)(0,0)
\put(106.06,259.94){\makebox(0,0)[bc]{0.60}}
\put(169.87,259.94){\makebox(0,0)[bc]{0.65}}
\put(162.63,50.13){\makebox(0,0)[lc]{-0.12}}
\put(162.63,84.18){\makebox(0,0)[lc]{-0.10}}
\put(162.63,118.23){\makebox(0,0)[lc]{-0.08}}
\put(162.63,186.32){\makebox(0,0)[lc]{-0.04}}
\put(162.63,220.37){\makebox(0,0)[lc]{-0.02}}
\put(0.00,259.94){\makebox(0,0)[bl]{Re(E)}}
\put(162.63,4.01){\makebox(0,0)[bl]{Im(E)}}
\put(190.44,96.66){\makebox(0,0){$\bullet$}}
\put(167.12,155.88){\makebox(0,0){$\bullet$}}
\put(145.21,210.79){\makebox(0,0){$\bullet$}}
\put(136.90,247.46){\makebox(0,0){$\bullet$}}
\put(121.19,254.42){\makebox(0,0){$\bullet$}}
\put(92.86,254.42){\makebox(0,0){$\bullet$}}
\put(62.99,254.42){\makebox(0,0){$\bullet$}}
\put(43.91,254.42){\makebox(0,0){$\bullet$}}
\put(30.10,254.42){\makebox(0,0){$\bullet$}}
\scriptsize
\put(135.88,254.42){\makebox(0,0){$\odot$}}
\put(150.01,254.42){\makebox(0,0){$\odot$}}
\put(155.14,254.42){\makebox(0,0){$\odot$}}
\put(157.06,254.42){\makebox(0,0){$\odot$}}
\put(156.72,254.42){\makebox(0,0){$\odot$}}
\put(155.98,254.42){\makebox(0,0){$\odot$}}
\put(154.66,254.42){\makebox(0,0){$\odot$}}
\put(152.46,254.42){\makebox(0,0){$\odot$}}
\put(148.54,254.42){\makebox(0,0){$\odot$}}
\footnotesize
\put(194,96.66){\makebox(0,0)[bl]{1.0}}
\put(170,155.88){\makebox(0,0)[bl]{1.1}}
\put(142,210.79){\makebox(0,0)[tr]{1.2}}
\put(103,181){\vector(1,2){32}}
\put(103,178){\makebox(0,0)[tc]{1.24}}
\put(121,251){\makebox(0,0)[tc]{1.25}}
\put(93,251){\makebox(0,0)[tc]{1.3}}
\put(63,251){\makebox(0,0)[tc]{1.4}}
\put(44,251){\makebox(0,0)[tc]{1.5}}
\put(30,251){\makebox(0,0)[tr]{1.6}}
\scriptsize
\put(5,75){\makebox(0,0)[bl]{
\begin{picture}(141.73,66.69)(0.00,0.00)
\put(51.51,41.90){\makebox(0,0)[bc]{0.63}}
\put(101.59,41.90){\makebox(0,0)[bc]{0.64}}
\put(0.00,41.90){\makebox(0,0)[bl]{Re(E)}}
\put(115.02,4.01){\makebox(0,0)[bl]{Im(E)}}
\put(39,16.2){\vector(1,0){10}}
\put(49,22.2){\vector(-1,0){10}}
\put(22.16,5){\makebox(0,0){$\bullet$}}
\put(18.29,19.27){\makebox(0,0){$\bullet$}}
\put(73.72,19.27){\makebox(0,0){$\bullet$}}
\put(93.85,19.27){\makebox(0,0){$\bullet$}}
\put(101.39,19.27){\makebox(0,0){$\bullet$}}
\put(100.07,19.27){\makebox(0,0){$\bullet$}}
\put(97.15,19.27){\makebox(0,0){$\bullet$}}
\put(92.00,19.27){\makebox(0,0){$\bullet$}}
\put(83.37,19.27){\makebox(0,0){$\bullet$}}
\put(67.98,19.27){\makebox(0,0){$\bullet$}}
\put(24,6){\makebox(0,0)[bl]{1.24}}
\put(18.3,22){\makebox(0,0)[bc]{1.5}}
\put(73.7,29){\vector(0,-1){7}}
\put(73.7,31){\makebox(0,0)[bc]{1.4}}
\put(93.9,29){\vector(0,-1){7}}
\put(93.9,31){\makebox(0,0)[bc]{1.35}}
\put(109,22){\makebox(0,0)[br]{1.3}}
\put(104,6){\vector(-1,3){4}}
\put(109,6){\makebox(0,0)[tr]{1.29}}
\put(97.2,15){\vector(0,1){3}}
\put(93.2,13.5){\makebox(0,0)[tc]{1.28}}
\put(73,6){\vector(4,3){16}}
\put(76,6){\makebox(0,0)[tc]{1.27}}
\put(64.4,6){\vector(4,3){16}}
\put(63,8){\makebox(0,0)[tr]{1.26}}
\put(68,16){\makebox(0,0)[tr]{1.25}}
\end{picture}
}}
\normalsize
\put(5,5){\makebox(0,0)[bl]{\bm{K\pi\rightarrow K\pi}}}
\put(40,145){\makebox(0,0)[bl]{\bm{E} {\bf plane}}}
\end{picture}
\end{center}
\normalsize
\caption[]{Hypothetical movement of the $K_{0}^{\ast}(730)$ pole in
$K\pi$ $S$-wave scattering as a function of the coupling constant.
The two branches on the imaginary $k$ axis are discussed in the text.
In the $E=\sqrt{s}$ plane these two branches come out on the real axis
below threshold.
The poles of the upper branch are shown as open circles in the main
figure, and as closed circles in the inset.
Units are in GeVs.}
\label{kappapole}
\end{figure}

In Fig.~(\ref{kappapole}) we study the hypothetical pole
positions of the $K_{0}^{\ast}(730)$ pole in $K\pi$
$S$-wave scattering.
The physical value of the coupling constant equals 0.75,
which is not shown in Fig.~(\ref{kappapole}).
A figure for smaller values of the coupling constants can be
found in Ref.~\cite{HEPEX0106077}.
The physical pole in $K\pi$ isodoublet $S$-wave scattering,
related to experiment \cite{HEPEX0204018,HEPEX0110052},
comes out at $727-i263$ MeV in Ref.~\cite{ZPC30p615}.
Here we concentrate on the threshold behaviour of the
hypothetical pole movements in the complex $k$ and $\sqrt{s}$
planes.
Until they meet on the axis, which is for a value of the
coupling constant slightly larger than 1.24, we have only
depicted the right-hand branch.

In the left figure we observe how the poles arrive on the imaginary
$k$ axis, and then continue to move along that axis.
One of the poles moves upwards, initially describing a virtual
bound state, and crossing the real $k$ axis for a value of the
coupling constant slightly larger than 1.30.
The other pole moves downwards, remaining a virtual bound state
for further increasing values of the coupling constant.

In the right figure the same pole has been depicted in the
$\sqrt{s}$ plane.
Here, the situation is more confusing, since the pole positions
are in the same interval of energies.
The pole which corresponds to the one moving downwards
along the imaginary $k$ axis moves to the left on the real
$\sqrt{s}$ axis.
Its positions as a function of the coupling constant are
indicated by solid circles.
The pole which moves upwards along the imaginary $k$ axis
initially moves towards threshold and then turns back,
following the former pole, but in a different Riemann sheet.
The positions of the latter pole are indicated by open circles.
In the inset we try to better clarify its motion.
Notice that, since we took 0.14 GeV and 0.50 GeV for
the pion and the Kaon mass, respectively, we end up
with a threshold at 0.64 GeV.

It is interesting to notice that in a recent work of M. Boglione and
M.R. Pennington \cite{HEPPH0203149} also a zero-width state is found
below the $K\pi$ threshold in $S$-wave scattering.
Here, we obtain such a state for \em unphysical \em \/values of the coupling.

\begin{figure}[ht]
\normalsize
\begin{center}
\begin{picture}(283.46,180.08)(0.00,0.00)
\put(71.03,146.12){\makebox(0,0)[bc]{0.70}}
\put(125.46,146.12){\makebox(0,0)[bc]{0.80}}
\put(179.90,146.12){\makebox(0,0)[bc]{0.90}}
\put(224.34,146.12){\makebox(0,0)[bl]{1.00 GeV}}
\put(239.86,55.06){\makebox(0,0)[lc]{-2.0 GeV}}
\put(239.86,97.83){\makebox(0,0)[lc]{-1.0 GeV}}
\put(0.00,146.12){\makebox(0,0)[bl]{Re(E)}}
\put(239.86,4.01){\makebox(0,0)[bl]{Im(E)}}
\put(132.10,140.60){\makebox(0,0){$\bullet$}}
\put(167.43,140.60){\makebox(0,0){$\bullet$}}
\put(222.60,140.60){\makebox(0,0){$\bullet$}}
\put(228.10,140.60){\makebox(0,0){$\bullet$}}
\put(234.30,140.60){\makebox(0,0){$\bullet$}}
\put(233.39,140.60){\makebox(0,0){$\bullet$}}
\put(229.44,140.60){\makebox(0,0){$\bullet$}}
\put(223.56,140.60){\makebox(0,0){$\bullet$}}
\put(208.43,140.60){\makebox(0,0){$\bullet$}}
\put(30.65,19.70){\makebox(0,0){$\bullet$}}
\put(66.78,67.54){\makebox(0,0){$\bullet$}}
\put(106.72,136.06){\makebox(0,0){$\bullet$}}
\footnotesize
\put(144,136){\makebox(0,0)[rt]{0.51}}
\put(179,136){\makebox(0,0)[rt]{0.52}}
\put(206,136){\makebox(0,0)[rt]{0.9}}
\put(36,20){\makebox(0,0)[lc]{0.02}}
\put(73,67){\makebox(0,0)[lc]{0.1}}
\put(102,136){\makebox(0,0)[rt]{0.5}}
\put(10,125){\makebox(0,0)[lt]{$KK\rightarrow KK$}}
\scriptsize
\put(75,25){\makebox(0,0)[bl]{
\begin{picture}(170,20)(0,138.78)
\put(45.03,155){\makebox(0,0)[bc]{0.98}}
\put(90.06,155){\makebox(0,0)[bc]{0.99}}
\put(135.03,155){\makebox(0,0)[bc]{1.00}}
\put(0.00,153.26){\makebox(0,0)[bl]{Re(E)}}
\put(37.94,149.34){\makebox(0,0){$\bullet$}}
\put(83.46,149.34){\makebox(0,0){$\bullet$}}
\put(134.76,149.34){\makebox(0,0){$\bullet$}}
\put(127.26,149.34){\makebox(0,0){$\bullet$}}
\put(94.58,149.34){\makebox(0,0){$\bullet$}}
\put(45.91,149.34){\makebox(0,0){$\bullet$}}
\footnotesize
\put(37.94,145){\makebox(0,0)[tc]{0.58}}
\put(83.46,145){\makebox(0,0)[tc]{0.60}}
\put(134.76,145){\makebox(0,0)[tc]{0.66}}
\put(127.26,135){\makebox(0,0)[tc]{0.70}}
\put(94.58,135){\makebox(0,0)[tc]{0.75}}
\put(45.91,135){\makebox(0,0)[tc]{0.80}}
\end{picture}}}
\normalsize
\end{picture}
\end{center}
\normalsize
\caption[]{Pole movement as a function of the coupling constant for
$KK$ $I=1$ $S$-wave scattering.
Some values of the coupling constant are indicated in the figure.
The six filled circles at the right end of the real axis correspond, from
left to right, to the values 0.58, 0.80, 0.60, 0.75, 0.70, and 0.66
for the model's coupling constant. This situation is magnified in the inset.}
\label{a0pole}
\end{figure}
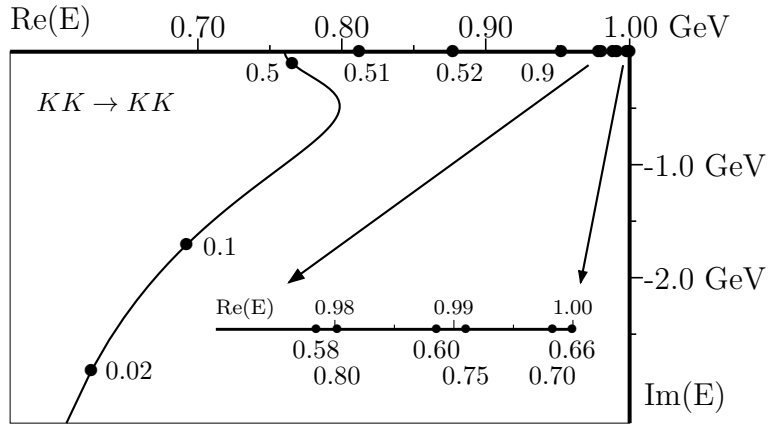

In Fig.~(\ref{a0pole}) we have depicted the movement of the
$a_ {0}(980)$ pole in $S$ wave $I=1$ $KK$ scattering
(threshold at 1.0 GeV) on the upwards-going branch.
One observes a very similar behaviour as in the case of $K\pi$
scattering, but with two important differences, to be described next.

The $K_{0}^{\ast}(730)$ poles meet on the real $\sqrt{s}$ axis
only 16 MeV below threshold (see Fig.~\ref{kappapole}),
and for a value of the coupling constant which is well above
the physical value of 0.75, whereas the $a_ {0}(980)$ poles meet
238 MeV below threshold, when the coupling constant
only equals 0.51.
At the physical value of the coupling constant, the $a_ {0}(980)$
pole is a real bound state some 9 MeV below threshold.

But there is yet another difference.
Whereas the $K\pi$ channel represents the lowest possible scattering
threshold for the $K_{0}^{\ast}(730)$ system,
$KK$ is not the lowest channel for the $a_ {0}(980)$.
In a more complete description, at least all pseudoscalar
meson-meson loops should be taken into account.
One of these is the $\eta\pi$ channel, which has a threshold
well below $KK$.
Consequently, upon including the $\eta\pi$ channel in the model, the pole
cannot remain on the real $\sqrt{s}$ axis, but has to acquire
an imaginary part in a similar way as shown in
Fig.~(\ref{Above}).
In Ref.~\cite{ZPC30p615} we obtained a resonance-like structure
in the $\eta\pi$ cross section, representing the physical
$a_{0}(980)$. The corresponding pole came out at $962-i28$ MeV.

For the $f_{0}(980)$ system the situation is very similar
to that of the $a_ {0}(980)$.
Assuming a pure $s\bar{s}$ quark content \cite{PLB521p15}, we obtain for the
variation of the corresponding pole in $KK$ $I=0$ $S$-wave scattering
a picture almost equal to the one shown in Fig.~(\ref{a0pole}).
However, only in lowest order the $KK$ channel could be considered
the lowest threshold for the $f_{0}(980)$ system.
In reality $s\bar{s}$ also couples to the nonstrange quark-antiquark
isosinglet through $KK$, and hence to $\pi\pi$ \cite{NPB266p451}.
This coupling is nevertheless very weak, which implies that the
resulting pole does not move far away from the $KK$ bound state.
In Ref.~\cite{ZPC30p615} we obtained a resonance-like structure
in the $\pi\pi$ cross section representing the physical
$f_ {0}(980)$.
The corresponding pole came out at $994-i20$ MeV.

At lower energies, we found for the same cross section a pole
which is the equivalent of the $K_{0}^{\ast}(730)$ system, but now
in $\pi\pi$ isoscalar $S$-wave scattering.
This pole at $470-i208$ MeV may be associated with the
$\sigma$ meson, since it has the same quantum numbers
and lies in the ballpark of predicted pole positions
in $\sigma$ models (for a complete overview of $\sigma$ poles,
see Ref.~\cite{HEPPH0201006}).

We do not find any other relevant poles in the energy region
up to 1.0 GeV.

\section{Conclusions}

We have shown that the poles of the $a_{0}(980)$ and $f_{0}(980)$
belong to a nonet of scattering-matrix poles.
The lower-lying isoscalar pole and the isodoublet poles in the complex-energy
plane have real parts of 0.47 GeV and 0.73 GeV, respectively, and imaginary
parts of 0.21 GeV resp.\ 0.26 GeV.
Whether these poles represent real physical resonances
\cite{NPA688p823} is not so relevant here.
Important is that the $a_{0}(980)$ and $f_{0}(980)$ are well
classified within a nonet of scattering-matrix poles with very
specific characteristics, different from those of the poles stemming
from confinement, like the confinement-ground-state nonet of scalar mesons
$f_{0}(1370)$, $a_{0}(1450)$, $K_{0}^{\ast}(1430)$, and $f_{0}(1500)$.
The latter poles vary as a function of the coupling constant
exactly the way indicated in figure (\ref{Above}).
For vanishing coupling constant they end up on the real $\sqrt{s}$ axis
at the positions of the various ground-state eigenvalues
of the confinement spectrum, which are the light-flavour
$^{3}P_{0}$ states at 1.3 to 1.5 GeV
\cite{PRD61p014015,HEPLAT9805029,NPPS53p236}.

The low-lying $S$-wave poles related to the cross sections
in the $f_{0}(470)$, $K_{0}^{\ast}(730)$, $f_{0}(980)$, and $a_{0}(980)$
regions move to negative imaginary infinity in the $\sqrt{s}$ plane
for decreasing values of the coupling.
The pole positions for the physical value of the coupling are well
explained by their threshold behaviour.
Whether or not these poles have large imaginary parts, leading to large
widths and strong resonance distortion, depends in a subtle way on the
thresholds and couplings of the various relevant scattering channels
\cite{PLB462p14}.

As to the nature of the light scalar mesons, which has recently been discussed
in Refs.~\cite{HEPPH0204205,HEPPH0201171,NPA675p209c,PRD65p114011},
we can only remark that in a many-coupled-channel model each of the
channels contributes to the states under the resonance, not just one
specific channel.

Comparing to experiment cross sections and phase shifts that follow from
modelling strong interactions is a particularly useful strategy,
since then the model gains independence with respect to the criteria
for omitting experimental results from the spectroscopic tables
\cite{HEPPH0202157,EKlemptProtvino01}.
Moreover, this strategy has another advantage.
Phase shifts do not easily follow from confinement models,
nor does Nature provide a bound-state spectrum of mesons.
Confinement models and spectroscopic tables go well together,
but do not shed sufficient light on strong interactions.
Both can be avoided when following the strategy outlined above.
Nonetheless, it would be useful to have some coordination in
the availability of experimental results on meson-meson scattering.
\vspace{0.3cm}

{\bf Acknowledgement}:
We wish to thank Frieder Kleefeld for useful discussions on the distribution
of $S$-matrix poles in the complex $E$ and $k$ planes.

This work has been partly supported by the
{\it Fun\-da\-\c{c}\~{a}o para a Ci\^{e}ncia e a Tecnologia}
\/of the {\it Minist\'{e}rio da
Ci\^{e}ncia e da Tecnologia} \/of Portugal,
under contract numbers
POCTI/\-35304/\-FIS/\-2000,
and
CERN/\-P/\-FIS/\-40119/\-2000.

\end{document}